\documentclass{elsarticle}
\usepackage{natbib,hyperref}
\usepackage{url}
\usepackage{graphicx}
\usepackage{amsmath}

\def\sgr{SGR\,1806--20}

\def\chandra{\emph{Chandra}}
\def\xmm{\emph{XMM-Newton}}

\begin{document}

\bibliographystyle{model2-names}\biboptions{authoryear}

\newcommand\arcdeg{\mbox{$^\circ$}}%
\newcommand\arcmin{\mbox{$^\prime$}}%
\newcommand\arcsec{\mbox{$^{\prime\prime}$}}%
\newcommand\fd{\mbox{$.\!\!^{\mathrm d}$}}%
\newcommand\fh{\mbox{$.\!\!^{\mathrm h}$}}%
\newcommand\fm{\mbox{$.\!\!^{\mathrm m}$}}%
\newcommand\fs{\mbox{$.\!\!^{\mathrm s}$}}%
\newcommand\fdg{\mbox{$.\!\!^\circ$}}%
\newcommand\farcm{\mbox{$.\mkern-4mu^\prime$}}%
\newcommand\farcs{\mbox{$.\!\!^{\prime\prime}$}}%
\newcommand\degr{\arcdeg}%

\begin{frontmatter}

\title{Searching for small-scale diffuse emission around \sgr}

% \author{D. Vigan\`o\inst{1}, N. Rea\inst{1,2}, P. Esposito\inst{3}, G.~L.~Israel\inst{4}, A.~Tiengo\inst{3,5}, S.~Mereghetti\inst{3}, R.~Turolla\inst{6}, S.~Zane\inst{7}, L.~Stella\inst{8}, D.~G\"otz\inst{9}}%\altaffilmark{4,5,6,7}}
%R. Turolla\altaffilmark{3,4}, L. Stella\altaffilmark{5},  M. Murgia\altaffilmark{2}, S. Zane\altaffilmark{4}, G.L. Israel\altaffilmark{5}, D. G\"otz\altaffilmark{6}, ...\altaffilmark{7}}
% \institute{Institut de Ci\`encies de l'Espai (CSIC-IEEC), Campus UAB, Facultat de Ci\`encies, Torre C5-parell, E-08193 Barcelona, Spain.
% \and
% Astronomical Institute ``Anton Pannekoek'', University of Amsterdam, Postbus 94249, 1090GE Amsterdam, The Netherlands.
% \and
% INAF--Istituto di Astrofisica Spaziale e Fisica Cosmica Milano, via E. Bassini 15, I-20133 Milano, Italy.
% \and
% INAF Roma
% \and
% Pavia
% \and
% Padova
% \and
% Mullard Space Science Laboratory, University College of London, Holmbury St Mary, Dorking, Surrey, RH5 6NT,  UK
% }

\author[ice]{D.~Vigan\`o}
\author[api,ice]{N.~Rea}
\author[inafmi]{P.~Esposito}
\author[inafmi]{S.~Mereghetti}
\author[inafrm]{G.~L.~Israel}
\author[pv,inafmi,infnpv]{A.~Tiengo}
\author[pd,mullard]{R.~Turolla}
\author[mullard]{S.~Zane}
\author[inafrm]{L.~Stella}

\address[ice]{Institut de Ci\`encies de l'Espai (CSIC-IEEC), Campus UAB, Facultat de Ci\`encies, Torre C5-parell, E-08193 Barcelona, Spain.}
\address[api]{Anton Pannekoek Institut, University of Amsterdam, Postbus 94249, 1090GE Amsterdam, The Netherlands.}
\address[inafmi]{Istituto Nazionale di Astrofisica, IASF Milano, Via E. Bassini 15, I-20133 Milano, Italy.}
\address[pv]{Istituto Universitario di Studi Superiori, Piazza della Vittoria 15, I-27100 Pavia, Italy}
\address[infnpv]{Istituto Nazionale di Fisica Nucleare, Sezione di Pavia, via A. Bassi 6, I-27100 Pavia, Italy}
\address[pd]{Department of Physics and Astronomy, University of Padova, Via Marzolo 8, 35131 Padova, Italy}
\address[inafrm]{INAF, Osservatorio Astronomico di Roma, Via Frascati, 33, 00040 Monteporzio Catone (Roma), Italy}
\address[mullard]{Mullard Space Science Laboratory, University College London, Holmbury St Mary, Dorking, Surrey, RH5 6NT,  UK}

% \thispagestyle{empty}

%\abstract
%{Diffuse radio emission was detected around the soft gamma-ray repeater \sgr\, after its 2004 powerful giant flare.}
%{We study the possible extended X-ray emission at small scales around \sgr, in two observations by the High Resolution Camera Spectrometer (HRC-S) on board of the \emph{Chandra X-ray Observatory}: in 2005, 115 days after the giant flare, and in 2013, during quiescence.}
%{We compare the radial profiles extracted from data images and PSF simulations, carefully considering various issues related with the uncertain calibration of the HRC PSF at sub-arcsec scales.}
%{We do not see statistically significant excesses pointing to an extended emission on scales of $\sim$ arcsec.}
%{To correctly analyze HRC images at sub-arcsecond scale, calibration issues need to be carefully considered. Only an improvement of the modeling of such uncertainties could lead to a definitive assessment on the possible \sgr X-ray diffuse emission.}

\begin{abstract}
Diffuse radio emission was detected around the soft gamma-ray repeater \sgr\, after its 2004 powerful giant flare. We study the possible extended X-ray emission at small scales around \sgr, in two observations by the High Resolution Camera Spectrometer (HRC-S) on board of the \emph{Chandra X-ray Observatory}: in 2005, 115 days after the giant flare, and in 2013, during quiescence. We compare the radial profiles extracted from data images and PSF simulations, carefully considering various issues related with the uncertain calibration of the HRC PSF at sub-arcsecond scales.
We do not see statistically significant excesses pointing to an extended emission on scales of arcseconds. As a consequence, \sgr\ is compatible with being point-like in X-rays, months after the giant flare, as well as in quiescence.
\end{abstract}


\end{frontmatter}

% \maketitle

\section{Introduction}

Magnetars are highly magnetized neutron stars \citep{duncan92,thompson95,thompson96} which share a number of common properties, including long spin periods ($P\sim 0.3$--12~s), large spin-down rates $\dot P\sim 10^{-14}$--$10^{-10}$~s~s$^{-1}$), relatively bright and variable persistent luminosities ($L_{\mathrm{X}}\sim 10^{31-35}$~erg~s$^{-1}$), and the emission of powerful bursts (see e.g. \citealt{mereghetti08,rea11} for recent reviews).
                                                                                                                                                                                                                                                                                                                                                                                                                                                                                                                                                                                                                                                                                                                                                                                                                                                                                                                                                                                            
Magnetar bursting/flaring events can be classified as short X-ray bursts (lasting $t < 0.1$\,s, X-ray luminosities $L\sim 10^{40-41}$~erg~s$^{-1}$), intermediate bursts ($t\sim1-60\,$s, $L\sim 10^{41-43}$~erg~s$^{-1}$), and the very energetic giant flares ($L\sim 10^{44-47}$~erg~s$^{-1}$), which have been detected only three times \citep{mazets79,hurley99,hurley05}. The 2004 December 27 giant flare from \sgr\, was exceptionally bright, with an initial hard spike lasting $\sim$0.2~s followed by a $\sim$500~s long pulsating tail \citep{hurley05,mereghetti05,palmer05}. The isotropic luminosity above 50\,keV was $\sim 2.3 \times 10^{47}$ erg s$^{-1}$ (for a distance of $\sim 9$ kpc; \citealt{tendulkar12} and references therein). Following this event a moving asymmetric nebula was discovered at radio frequencies \citep{cameron05,gaensler05,taylor05,gelfand05,fender06}. The extremely accurate localization ($\sim 0.1$arcsec) obtained with the radio data made it possible the identification of a variable infrared counterpart \citep{kosugi05,israel05} to \sgr. Well after the giant flare and a strong bursting episode in 2006, \cite{svirski11} detected X-ray diffuse emission around \sgr\, in quiescence, with a size of a few arcminutes, interpreted as dust scattering echo.

We report here on the study of the high angular resolution X-ray data of the \chandra\ HRC-S camera, taken a few months after the giant flare (2005) and in quiescence (2013), in order to investigate a possible small-scale X-ray diffuse emission around \sgr.

\begin{table*}
\begin{center}
\caption{Positions of the \sgr\ centroids, as given by X-rays, with errors at 1$\sigma$, and IR analyses. For the former, the positions are obtained by the {\tt celldetect} tool run on an image extracted with {\tt fluximage}, with {\tt binsize=0.5}.}
\label{tab:positions}
\begin{tabular}{l c c}
\hline 
\hline 
Method			& RA 	& Dec \\
\hline 
X-ray 2005		& $\rm 18^h08^m39\fs359~(\pm0\fs0005)$ & $\rm -20 \degr 24' 40\farcs 00~(\pm0\farcs007)$ \\
X-ray burst 2005	& $\rm 18^h08^m39\fs337~(\pm0\fs003)$ & $\rm -20 \degr 24' 40\farcs 19~(\pm0\farcs041)$ \\
X-ray 2013 		& $\rm 18^h08^m39\fs366~(\pm0\fs0009)$ & $\rm -20 \degr 24' 40\farcs 37~(\pm0\farcs013)$ \\
IR \cite{kosugi05}	& $\rm 18^h08^m39\fs329$ & $\rm -20\degr24'39\farcs94$ \\
IR \cite{israel05}	& $\rm 18^h08^m39\fs337$ & $\rm -20\degr24'39\farcs85$ \\
\hline 
\hline
\end{tabular}

\end{center}

\end{table*} 
% 
% \begin{figure*}
% \includegraphics[width=.5\textwidth]{images/regions_2005.eps}
% \includegraphics[width=.5\textwidth]{images/regions_2013.eps}
% \caption{Positions in the X-ray image (left: 2005, right: 2013). The {\tt celldetect} position from the total events (green) and from 2005 bursts events (red) are plotted with the 1$\sigma$ uncertainty; the blue circle, whose size is arbitrary, marks the IR position.}
% \label{fig:positions}
% \end{figure*}

\section{Chandra HRC-S data analysis.}\label{obs}

The High Resolution Spectroscopy camera (\mbox{HRC-S}; \citealt{murray00}) on board \chandra\ is a multichannel plate detector sensitive to X-rays over the 0.1--10 keV energy range. The instrument has a 0.13$''$ pixel size, while essentially no energy information on the detected photons is available.\footnote{\url{http://cxc.harvard.edu/proposer/POG/html/chap7.html}  provides an updated description of HRC and related issues.} Here we analyze two observations, taken in timing mode, for which the instrument time resolution is 16~$\mu$s. The reprocessing, reduction and analysis of data were performed using CIAO 4.5 and the corresponding calibration libraries (CALDB 4.5.8). We reprocessed data applying the subpixel reconstruction by randomization ({\tt pix\_adj=randomize} in the {\tt chandra\_repro} routine), and extracted images and exposure maps with the {\tt fluximage} tool.

In order to improve the signal-to-noise (S/N) ratio, we considered only the channels {\tt PI} $\in [48,293]$.\footnote{\url{http://cxc.harvard.edu/ciao/threads/hrci_bg_spectra} This filter applies to the HRC Imaging (HRC-I) and HRC-S in timing mode.} This allowed us to reduce the instrumental background by about $25\%$, losing only $\sim 0.1\%$ of the source counts within the extraction region.

An additional filtering would consist in removing all the events with {\tt AMP\_SF}=3, in order to reduce the known asymmetric artifact at $\sim 0.6-0.8''$.\footnote{\url{http://cxc.harvard.edu/ciao/caveats/psf_artifact.html} and \url{http://hea-www.harvard.edu/~juda/memos/HEAD2010/} \url{HEAD2010_poster.html}} The latter is caused by an HRC hardware problem (the ringing of the amplifiers), which is only partly corrected by the processing of the raw data ({\tt hrc\_process\_events} tool). In our cases, such filter reduces the background by $\sim 25\%$, but also removes $\sim 3\%$ of the source counts. We checked that including or not such filter produces essentially identical images, so we did not apply it.

The net exposure time changes by maximum $1.5\%$ in different regions, and the maximum background relative variations in our observations are estimated to be less than $5\%$, across the field of view.

\begin{figure}
\centering
\includegraphics[width=.45\textwidth]{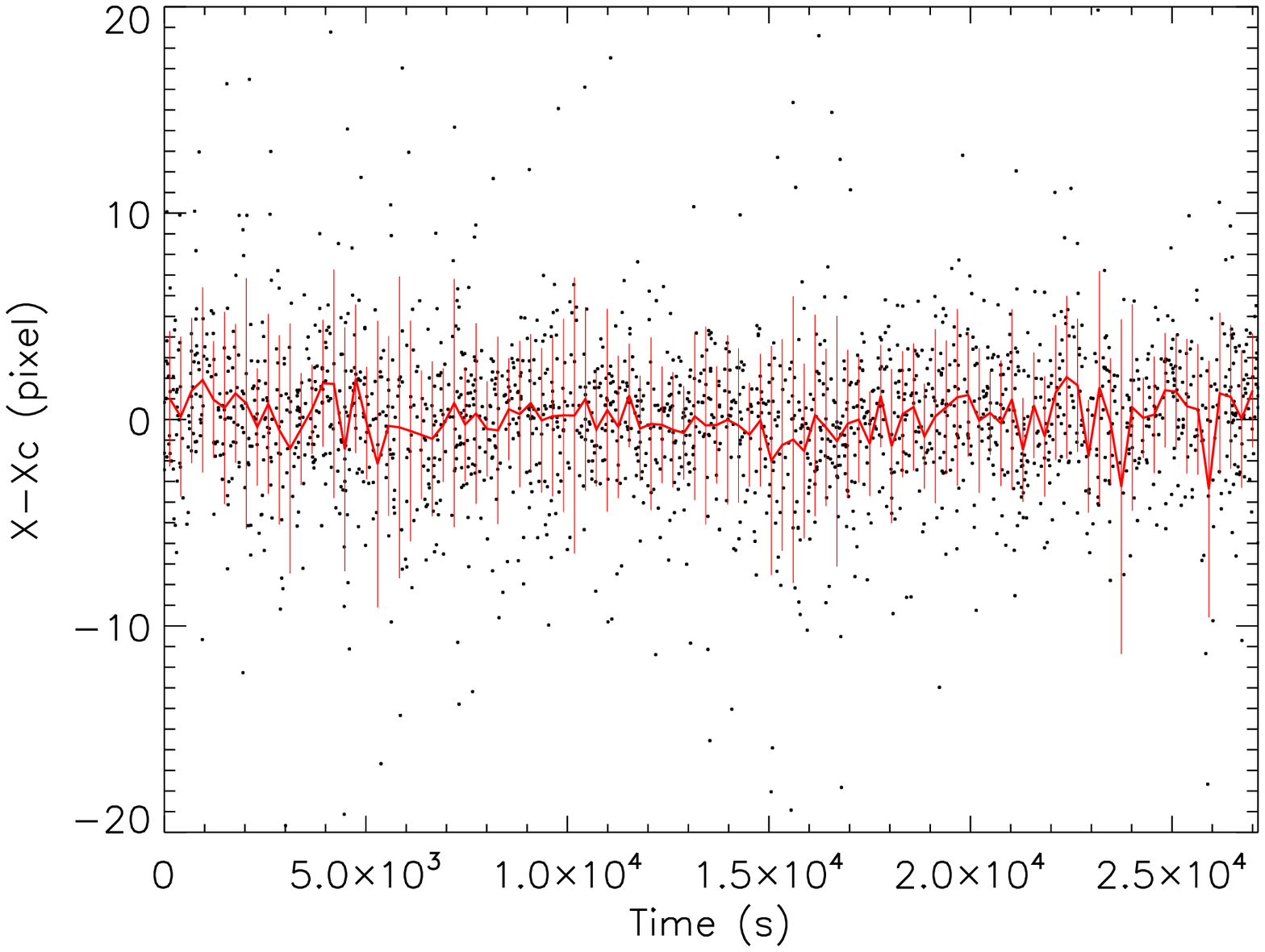}
\includegraphics[width=.45\textwidth]{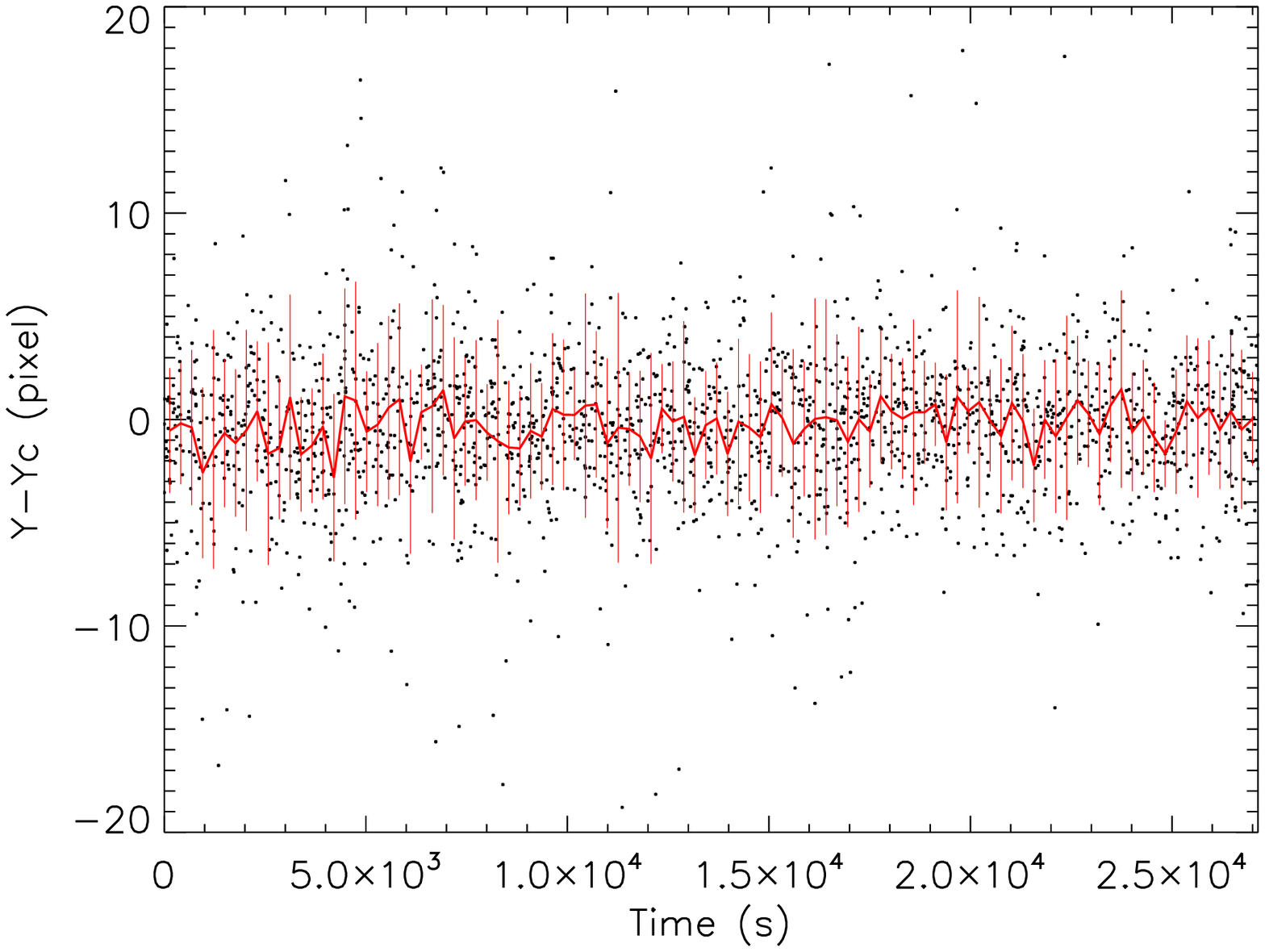}
\caption{Mean positions of X-ray events in detector coordinates (X and Y), as a function of time, indicated by black dots, and mean position (with the relative error) associated to each of the 100 time bins (red points and bars). Each time bin is $271$ s long, and contains, on average, 20 events.}
\label{fig:dithering} 
\end{figure}

% \begin{figure}
% \centering
% \includegraphics[width=.45\textwidth]{images/2005_spectra.eps}
% \caption{Fiducial, hard and soft spectra employed in the PSF simulations for the 2005 observation.}
% \label{fig:2005_spectra} 
% \end{figure}

\begin{table}
\begin{center}
\caption{Parameters of the spectral model used to simulate the PSF (BB+PL for the 2005 observation, and PL spectrum for the 2013 observation).}
\label{tab:spectra}
\begin{tabular}{l c c c c c}
\hline 
\hline 
Model			& $N_{\rm H}$ & $kT_{bb}$ & $R_{bb}$ & index & PL norm \\
				& [$10^{22}$cm$^{-2}$] & [keV] & $\left[\frac{{\rm km}^2}{(10 {\rm kpc})^2}\right]$ & & [$10^{-4}$] \\
\hline 
2005 & & & & & \\
\hline
Fiducial	&  5.87 & 0.93 & 1.61 & 0.66 & 7.5 \\
Soft		&  0.01 & 0.93 & 1.61 & 0.66 & 7.5 \\
Hard 	&  6.60 & 0.70 & 1.95 & 0.40 & 39 \\
\hline 
2013 & & & & & \\
\hline
Fiducial	&  6.50 & - & - & 2.50 & 93 \\
\hline 
\hline
\end{tabular}

\end{center}

\end{table}

The HRC-S observed \sgr~on 2005 April 22, 116 days after the giant flare, in a single, uninterrupted 27.1~ks timing mode exposure (27.0 ks after correcting for the dead time; ObsID 6251). After applying the filters described above, we extracted a total of 1927 counts from a circular region of $2''$ radius centered on the X-ray source. From a source-free, circular region of radius $\sim 20''$, far from \sgr, we estimate that the background contributes to $\sim 0.5\%$ of the total counts ($\sim$11 photons).

The inspection of the light curve binned at 0.1~s shows the presence of several short bursts. We removed the time intervals with the bursts from the event lists by applying intensity filters. This reduced the net exposure time by 1.9~s, during which $95 \pm 10$ counts attributable to the bursts from \sgr\ were detected.

With such time interval filter, assuming that the source is point-like and that $91\%$ of counts are enclosed within $2''$ (according to the simulated PSF below), the {\tt aprates} tool estimates the total counts of \sgr\ to be $2100 \pm 50$.

We repeated the same procedure and cross-checks for the 2013 May 15, 30~ks timing mode observation (ObsID 14884). We extracted a total of $660 \pm 26$ counts from a circular region of $2''$ radius centered on our X-ray source with an estimated background contribution of $\sim 3\%$ of the total counts ($\sim$ 20 photons). No bursts are seen in the light curve.

Since \sgr\ is the only point source detected in the observations, we use the {\tt celldetect} routine to obtain the best-fitting coordinates of its X-ray centroid, considering the exposure map and the image previously extracted. In 2005, additional information is provided by the counts associated with short bursts. Removing these events gives a negligible ($\sim 0.6$ mas) difference on the source best-fit position. On the other hand, when considering only the burst events, the X-ray position differs by $0\farcs38$ with respect to the position obtained considering the entire X-ray flux. Last, we note that the choice of a narrower {\tt binsize} parameter in {\tt fluximage} slightly changes the best-fit position, with further sub-pixel refinement ({\tt binsize}=0.1) providing only differences of the order $\sim 0\farcs07$.

The X-ray position between the two observations differ by $\sim 0\farcs38$, which is of the order of the Chandra absolute astrometric accuracy ($0\farcs6$ at $95\%$ c.l.). An infrared (IR) counterpart of \sgr\ was independently proposed by \cite{israel05} and \cite{kosugi05}. With respect to the position reported by the latter, our X-ray position has an offset of $\sim 0\farcs45$ (in 2005) and $\sim 0\farcs 70$ (in 2013). The recently estimated proper motion of $\sim 8.2$ mas yr$^{-1}$ (of the same order of the field velocity and the galactic rotation velocity, \citealt{tendulkar12}) translates into a negligible $\sim 0\farcs066$ difference expected between the source position in the 2005 and 2013 observation.

In Table~\ref{tab:positions} we report the positions we have found, and the IR positions reported in literature. The quoted statistical 1$\sigma$ uncertainties and the mutual discrepancies are both within the absolute astrometric accuracy of the pointing: $0\farcs4$ (at 68\% confidence level).\footnote{See \url{http://asc.harvard.edu/cal/ASPECT/celmon/}} The off-set of the source position with respect to the instrument aim point source position is $0\farcm28$ in 2005, and $0\farcm30$ in 2013.

% %%%%%%%%%%%%%%%%%%%%%%%%%%%%%%%%%%%%%%%%%%%%%%%%%%%%%%%%
\begin{figure*}
\centering
\includegraphics[width=.32\textwidth]{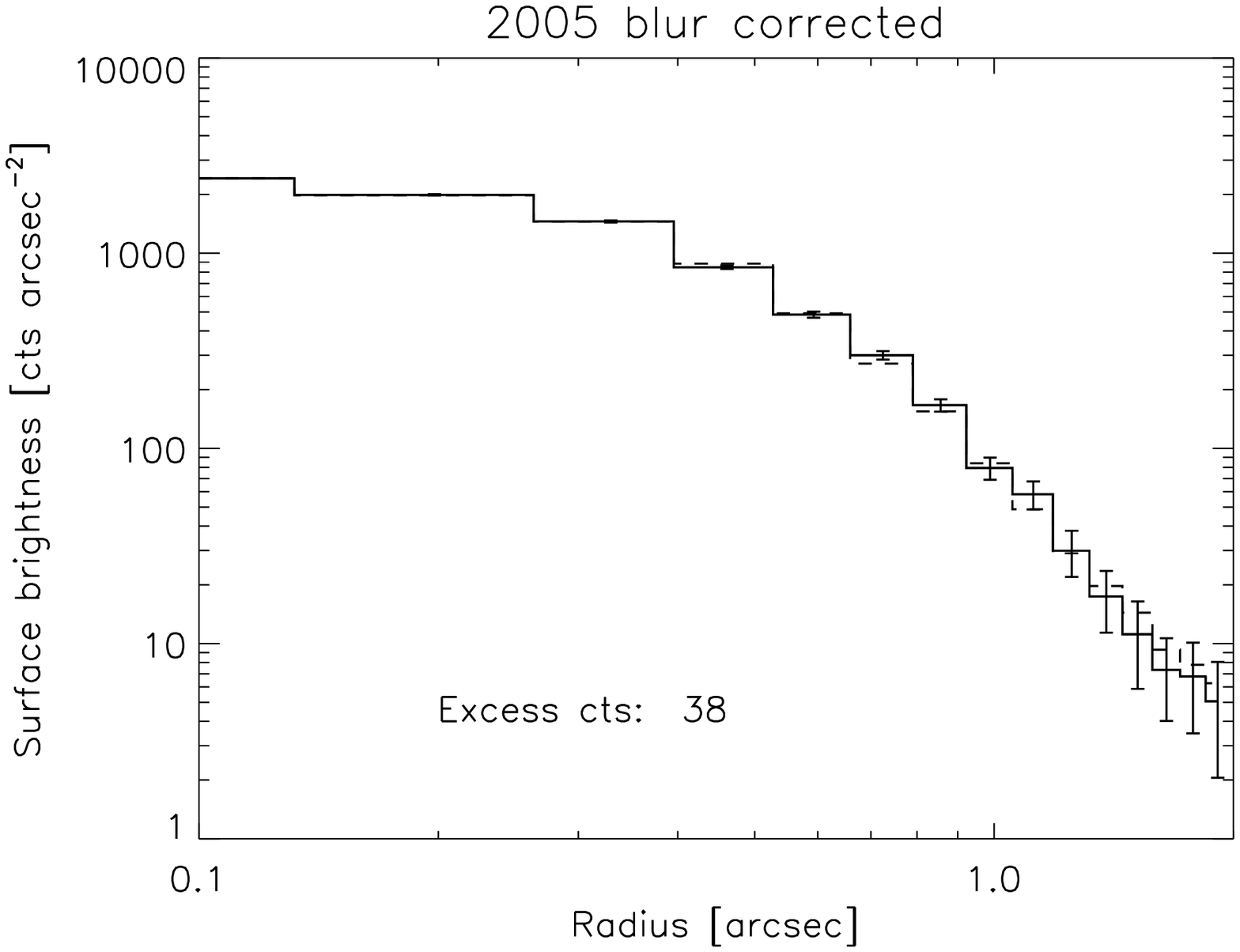}
\includegraphics[width=.32\textwidth]{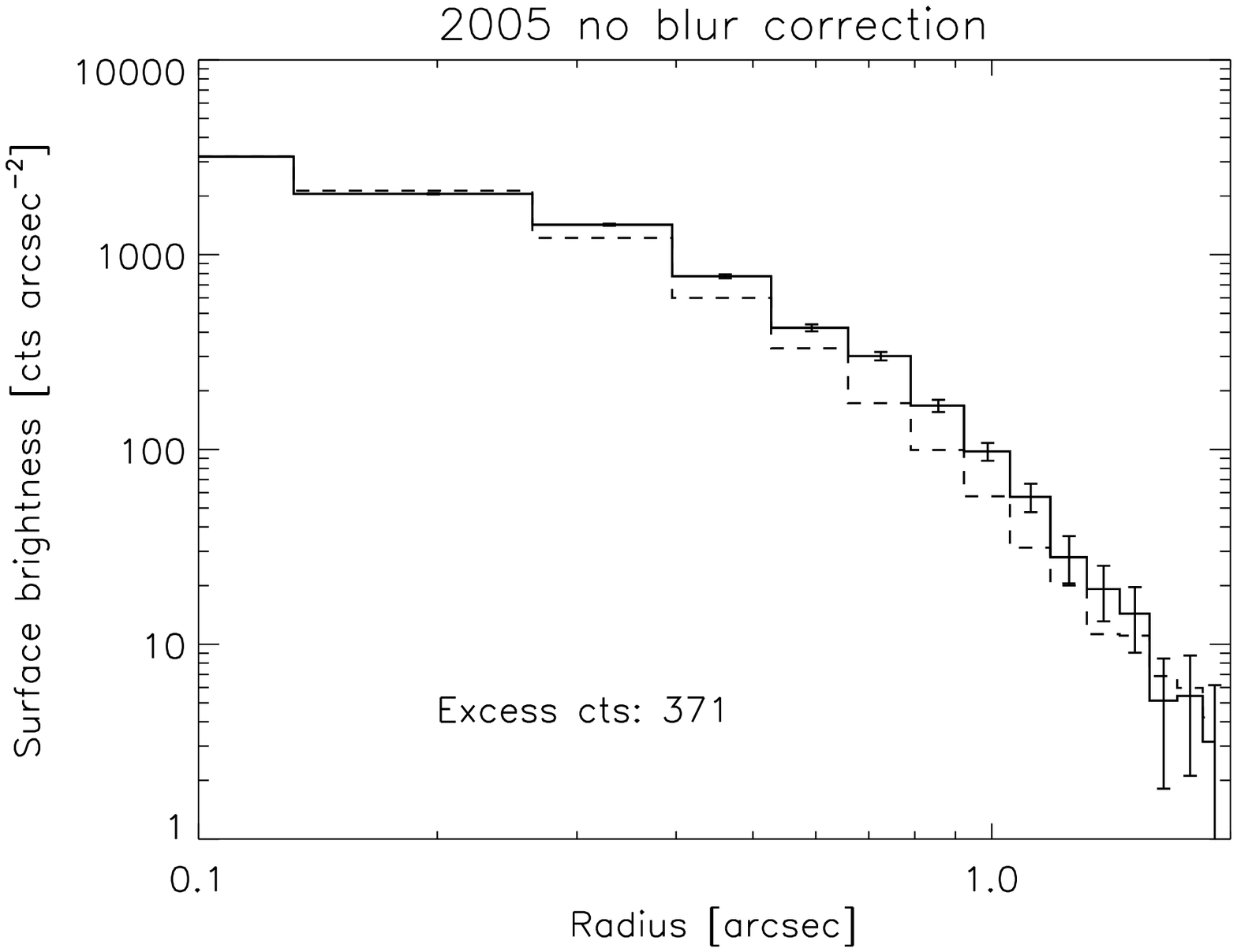}
\includegraphics[width=.32\textwidth]{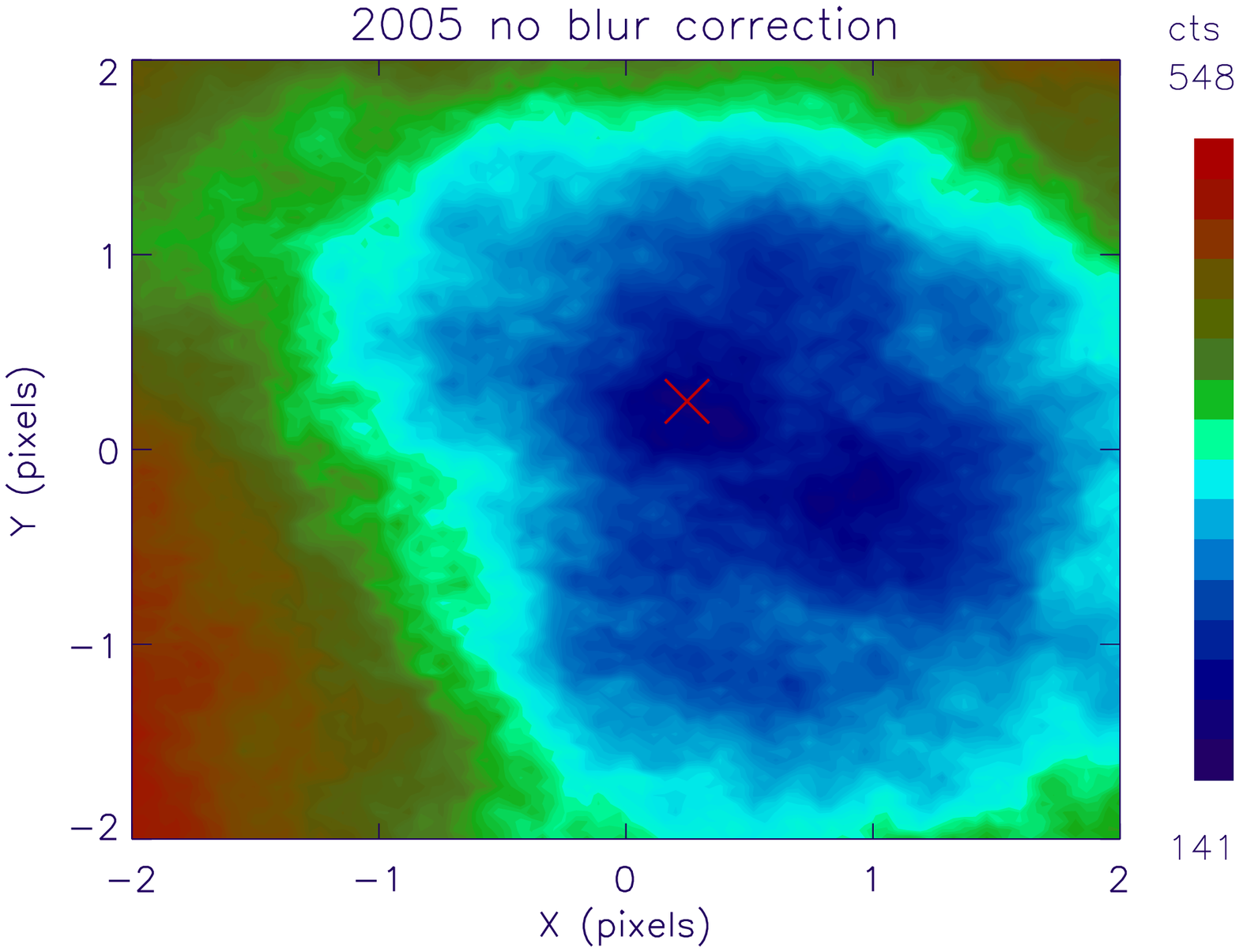}
\caption{Comparison between the radial profiles of 2005 data (solid line) and simulated PSF (dashes), for the fiducial spectra, with (left) and without (center) the blurring correction. Right panel: contour of the root mean square difference of counts, as a function of the assumed position of the center of the annuli used for the extraction of the radial profile in data. The position is in units of pixel, relative to the fiducial X-ray centroid position given by {\tt celldetect}.}
\label{fig:excess_contour_2005}
\end{figure*}

\begin{figure*}
\centering
\includegraphics[width=.32\textwidth]{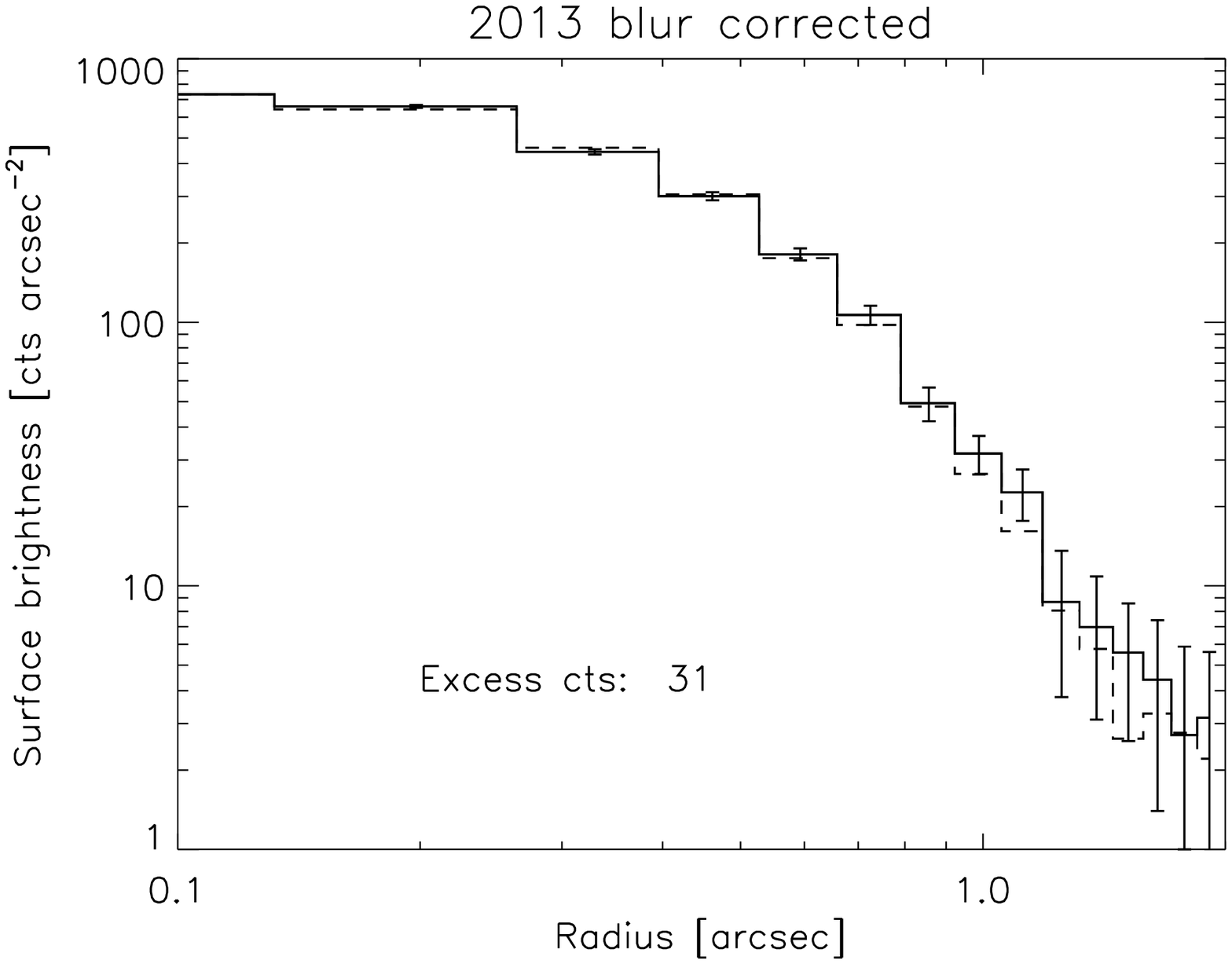}
\includegraphics[width=.32\textwidth]{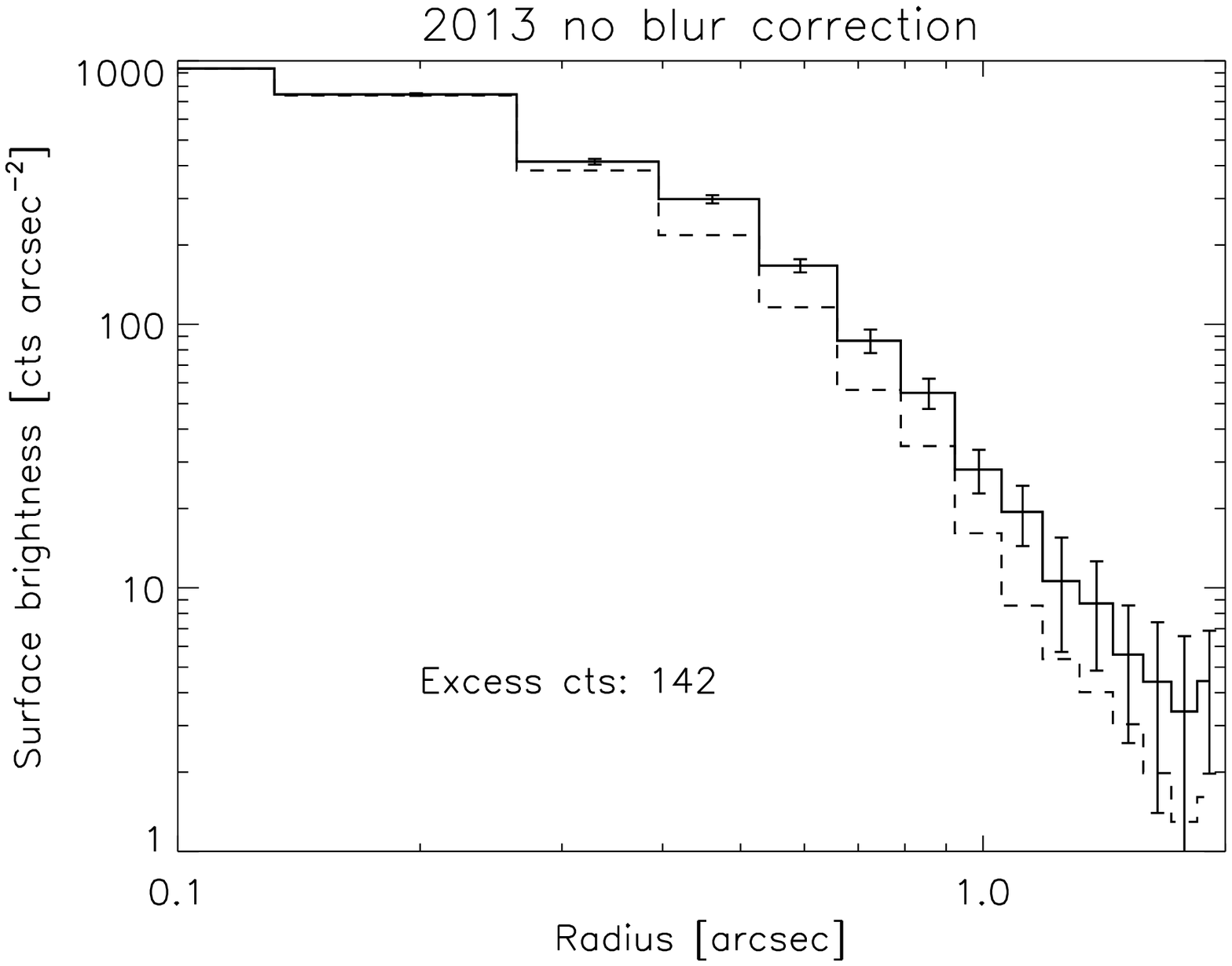}
\includegraphics[width=.32\textwidth]{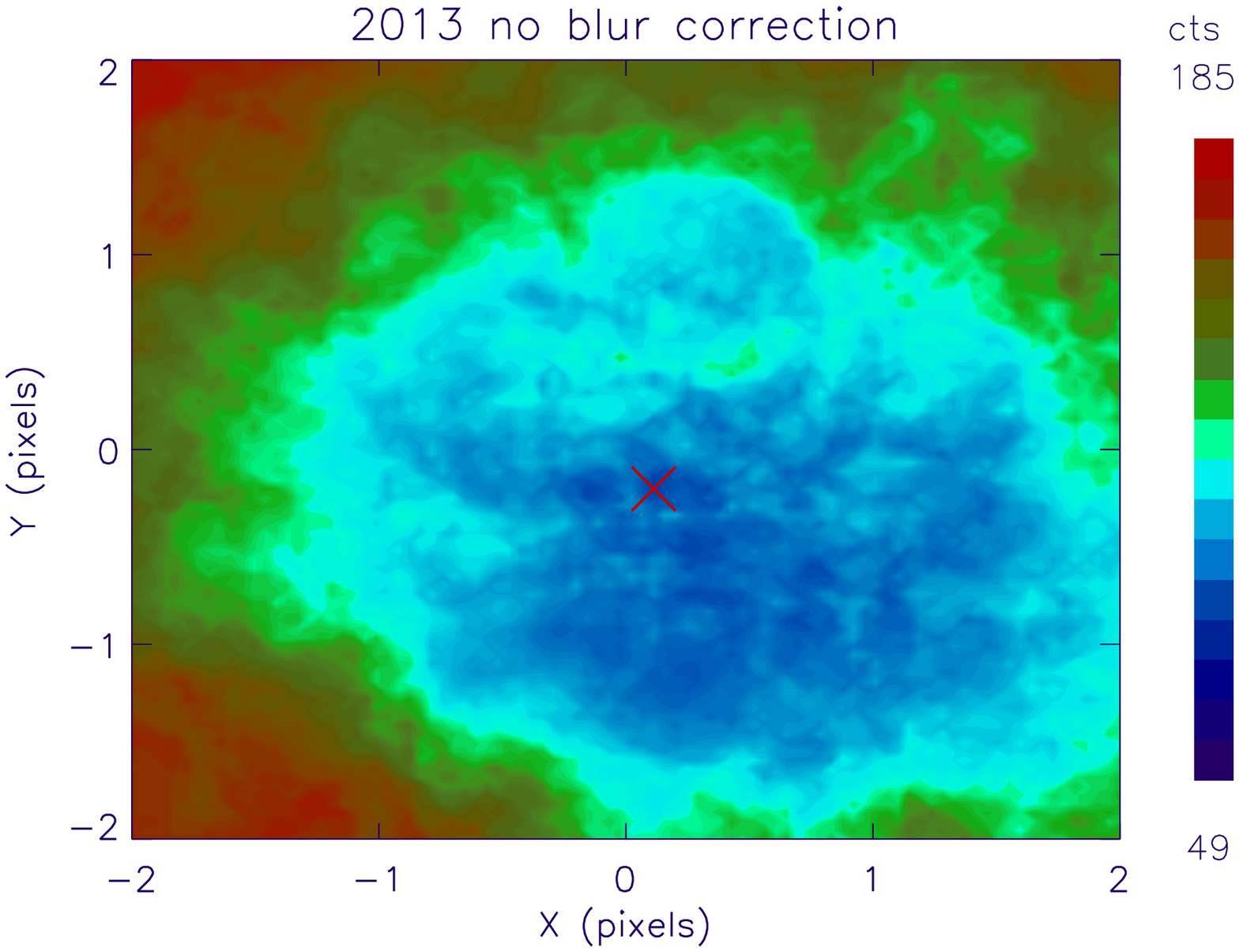}
\caption{Same as Fig.~\ref{fig:excess_contour_2005}, for the 2013 data.}
\label{fig:excess_contour_2013}
\end{figure*}

% \begin{figure*}
% \centering
% \includegraphics[width=.45\textwidth]{images/2005_excess_marx_blur.eps}
% \caption{Same as Fig.~\ref{fig:excess_contour_2005}, for the 2005 data, including the blurring correction.}
% \label{fig:excess_contour_2005_blur}
% \end{figure*}
% 
% \begin{figure*}
% \centering
% \includegraphics[width=.45\textwidth]{images/2013_profiles_marx_blur.eps}
% \caption{Same as Fig.~\ref{fig:excess_contour_2013}, for the 2013 data, including the blurring correction.}
% \label{fig:excess_contour_2013_blur}
% \end{figure*}

% \begin{figure*}
% \centering
% 
% \includegraphics[width=.45\textwidth]{images/1223_excess_marx.eps}
% \includegraphics[width=.45\textwidth]{images/1223_profiles_marx.eps}\\
% \includegraphics[width=.45\textwidth]{images/1223_excess_marx_blur.eps}
% \includegraphics[width=.45\textwidth]{images/1223_profiles_marx_blur.eps}
% \caption{Top: same as Fig.~\ref{fig:excess_contour_2005}, for the RBS 1223 data, accounting (top) or not (bottom) for the blurring correction.}
% \label{fig:profile_1223} 
% \end{figure*}

\section{Extended or point-like emission?}

\subsection{Simulated Point Spread Function}

To investigate the possible presence of extended emission around \sgr, we extracted the radial profiles from the images, and compared them to the Point Spread Function (PSF) profile expected from a point source, which was simulated using the Chandra Ray Tracer\footnote{See \url{http://cxc.harvard.edu/chart/index.html}} (ChaRT, March 2014 online version; \citealt{carter03}) and Model of AXAF Response to X-rays\footnote{See \url{http://space.mit.edu/cxc/marx/index.html}} (MARX v5.0.0-0) software packages.\footnote{We stress that, compared with initial results obtained with the 2012 version, the ChaRT/MARX simulations provide now significantly larger PSFs, especially at high energy.} For each observation, we used the satellite aspect (rolling angle and pointing) and the fiducial position derived from the data analysis, both for the construction of the PSF and the extraction of the radial profile. We checked that using other positions within the pointing accuracy error yields statistically negligible differences in the radial profiles.

There are some caveats regarding the modeling of the PSF of HRC on sub-arcesond scales.\footnote{See \url{http://cxc.harvard.edu/chart/caveats.html}} In particular, the residual errors in the position reconstruction cause a blur of the image over scales of $\sim 0\farcs2$ (for on-axis sources). Such errors could be in principle modeled by the analysis of the event positions as a function of time.\footnote{See the case of the calibration source AR Lac, Fig. 1 at \url{http://cxc.harvard.edu/contrib/juda/memos/hrc_blur/} \url{hrc_blur_update.html}}. In Fig.~\ref{fig:dithering} we show the mean positions (in detector coordinates) averaged out over time bins of 271 s, for the 2005 observation. In our case, the statistic is not high enough to recognize the possible residual position dithering (few pixels).

On the other hand, the blurring effect can be approximately considered in the ChaRT/MARX simulations by setting the phenomenological parameter {\tt Aspect\_Blur} larger than the default value of $0\farcs07$. Later we set it to $0\farcs15$. Such value, recommended by the manual, results from several years of calibration test, and it is an effective, phenomenological way to account for the effect. According to the manual, this value could be actually slightly larger, up $0\farcs20$, if high energy photons (several keV) are considered, since they get more blurred, compared to softX-ray photons. Last, we verified that including or not the pixel randomization in the simulations ({\tt pix\_adj} parameter set to {\tt exact} or {\tt randomize} in MARX) does not cause any statistically significant difference in the simulated image.

% To check the robustness of this result, we compared the observed profile also with PSFs obtained with different methods: using the CIAO tool mkpsf\footnote{See \url{http://cxc.harvard.edu/ciao/ahelp/mkpsf.html}.} (with the latest HRC-S calibration PSF libraries) and empirically, based on a HRC-S observation of the point source PSR\,J0218+4232 (ObsID 1853; \citealt{kuiper02}). All these approaches gave essentially identical results. We then chose to continue our analysis only using the {\tt ChaRT/MARX} software, which is widely reputed to simulate the best available PSF for a point source.

For the ray-tracing simulation, given the absence of spectral information in the HRC data, we have to assume a spectrum for the source. For the 2005 observation, we take the closest available \xmm\ data, on 2005 March 07, which can be fit by an absorbed power law plus blackbody model, shown in Table~\ref{tab:spectra} as fiducial, which is compatible with what obtained by \cite{tiengo05} (second row of their Table 1).\footnote{The third row of their Table 1 is obtained by the fit of five different observations between 2003 and 2005, by forcing the values of $N_{\rm H}$ and the blackbody component to be the same, with the power law component free to vary. The obtained spectrum has the following parameters: $N_{\mathrm{H}}=6.6\times10^{22}$cm$^{-2}$, $\Gamma=1.4$ and $kT=0.7$ keV. Note that during the first year after the giant flare, the luminosity of \sgr\ monotonically decreased by a factor of $\sim$2 and the spectrum was softening \citep{woods06,emt07}. Since there is no reason to expect the temperature to be the same, we prefer to use the fit from the single observation (spectrum A).} In order to test different spectra, we employed one much harder and one much softer, shown in Table~\ref{tab:spectra}. For the fiducial spectrum of the 2013 observation, we use the spectral fit to Swift/XRT data taken on 2013 March 17 (ObsID: 00035315026, $630 \pm 25$ photons): an absorbed power law with the parameters shown in Table~\ref{tab:spectra}.

\subsection{High-resolution radial profiles.}

We compare the radial profiles of data and simulations, obtained by considering the counts in a series of 1~pixel-wide annuli, centered around the same position. We normalize the simulated profile by fixing the first bin (i.e., the counts within $0\farcs13$) to be the same as in the observed profile.

To build the simulations, we employed the fiducial X-ray position discussed above. For the data, as discussed above, we are not able to determine the exact position of the source with an accuracy better than $\sim 0\farcs4$. To this purpose, we move the annuli center across a grid of 100x100 positions in a box of 4x4 pixels (i.e., $0\farcs53\times0\farcs53$) around the fiducial centroid of emission. Then, for each position, we systematically compute the radial profiles obtained for different annuli centers, and evaluate the root mean square (rms) differences and the counts excess. In this way, we are able to study how the assumed centroid position affects the differences between the theoretical and observed radial profiles, and to find, eventually, the optimal position, i.e. the one which minimizes the root mean square differences between the observed and simulated radial profiles.

In Fig.~\ref{fig:excess_contour_2005}, we show the comparison between the observed (solid lines) and simulated (dashes) radial profiles, for the 2005 case, with ({\tt Aspect\_Blur}=0.15, left) or without the blur correction ({\tt Aspect\_Blur}=0.07, center), for the optimal position, together with the total number of positive counts in excess (i.e., not taking into account the bins where data counts are less than in PSF). For the case without blur correction, we also show the 2D map of the excesses as a function of position (right panel). If the assumed position is strongly displaced from the optimal one, the profile becomes flatter, since the real peak of the emission is distributed among many radial bins, not only the central ones. Thus, a displaced center provides, by definition, large counts excesses (green and red colors).

% In the left panels, we represent the rms differences as a function of assumed position (difference increase from blue to red).
% The fiducial position lies in the center of the plot, while the optimal position is indicated with a cross. In the right panels, we show the comparison of radial profiles for data with the optimal position (solid lines), and simulated PSF (dashes).

In order to explore the effect of the spectrum on the simulated PSF, we have repeated the simulations for the three spectra of Table~\ref{tab:spectra}. The harder the spectrum, the broader the simulated PSF, and the smaller the rms differences. However, regardless of the spectrum, a few hundreds counts in excess (corresponding to a $\sim 15-25\%$ of the total counts) are seen in the optimal position only in the absence of the blur correction.

The same holds for the 2013 case, shown in Fig.~\ref{fig:excess_contour_2013}: only when the blurring correction is properly considered in the simulation ({\tt Aspect\_Blur}=$0\farcs15$), the expected and observed profiles are compatible (excess counts not statistically significant).
%, as shown in Figs.~\ref{fig:excess_contour_2005_blur} and \ref{fig:excess_contour_2013_blur}, for the 2005 and 2013 observations, respectively.

\subsection{Radial profile at intermediate scales.}

\begin{figure*}
\centering
\includegraphics[width=.45\textwidth]{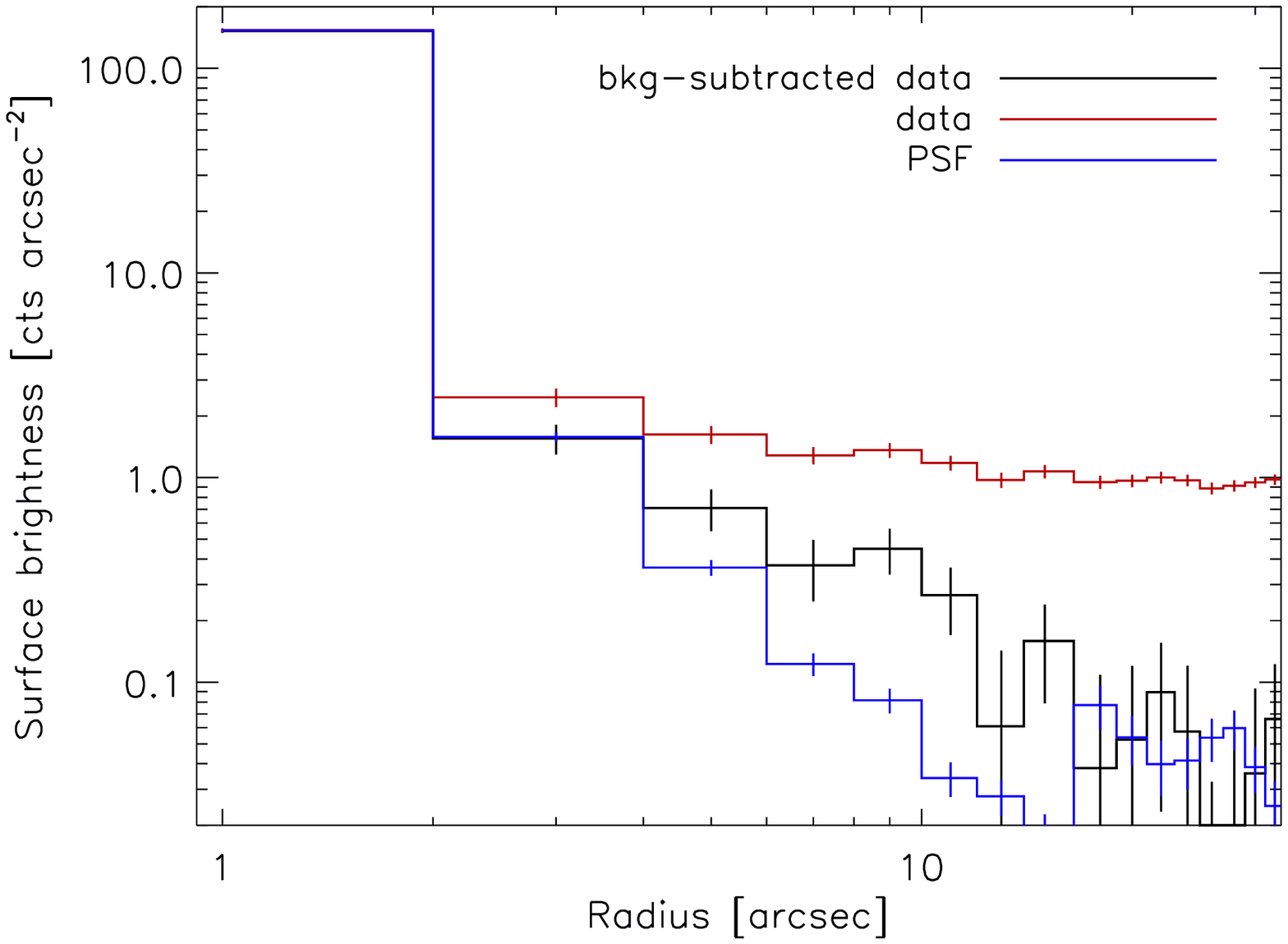}
\includegraphics[width=.45\textwidth]{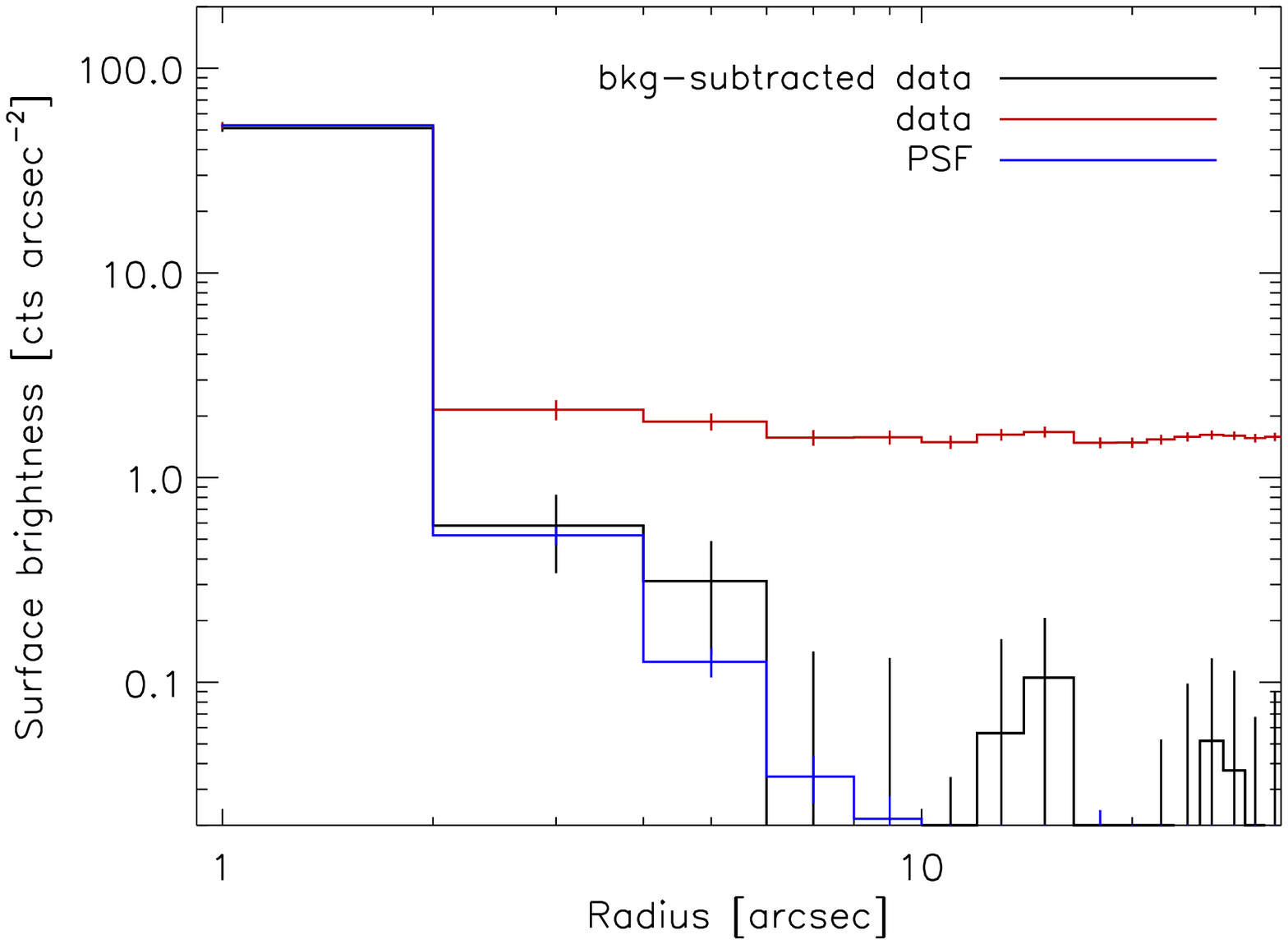}
\caption{Comparison at arcsecond scale of the radial profiles extracted from data (red), background-subtracted data (black), and PSF simulation (blue). Left: 2005 observation; right: 2013 observation.}
\label{fig:annuli2} 
\end{figure*}

We repeat the same procedure above at larger scales (several arcseconds), to obtain radial profiles of the counts comprised within a series of $2''$ annuli, up to $50''$ from the fiducial center. In Fig.~\ref{fig:annuli2} we plot the comparison, for the 2005 (left panel) and 2013 (right) observations, of the radial profiles as extracted from data (red), background-subtracted data (black), and PSF simulation (blue). We normalize the profiles to the number of counts within $2''$. In 2005, a significant excess of the background-subtracted data counts are evident between $4''$ and $12''$: in this annular region, there are $118 \pm 11$ counts in excess, which represent $22\%$ of the total counts in that region. Considering the circular region within $30''$, the excess counts are $204\pm 15$ ($2\%$ of the total counts in the same region). We have checked that this excess is present also when different spectra are used to simulate the PSF: for the hard spectrum of Table~\ref{tab:spectra}, the excess between  $4''$ and $12''$ is only slightly reduced: $112 \pm 11$ ($21\%$). We note also that, at such scales, the radial profiles are basically insensitive to the detailed position and the small-scale blurring: we have checked that the significance of the excess does not depend on the blurring correction.

On the other hand, in 2013 no clear excess is visible: the observed and simulated radial profiles seem compatible in the region within $6''$, where there are enough counts. However, the high background level would not allow to see a $\sim 20\%$ excess counts in the same region as 2005. This prevents us from comparing properly the diffuse emission in the two observations.

\section{Timing analysis.}

In the first \chandra\ observations, by using a $Z^2_1$
(Rayleigh test), we measured a period $P=7.5611\pm0.0017$~s, which is
in good agreement with the ephemeris by \citet{woods07}. The RMS
pulsed fraction was $10.2\pm1.3$~\% (uncertainties were
determined from Monte Carlo simulations). For the 2013 observation, we searched for the periodic signals between 7 and 8~s, but it could
not be found. This is not surprising, considering the low
pulsed fraction of \sgr, and the low counting statistics of the
2013 observation. To restrict the search to a narrower range,
%since no magnetar has ever been observed to spin up in the long-term,
we looked upon the last available values of the spin period of \sgr. The most recent observation of SGR 1806-20 where the period can be detected is an
%The last published measurement is $P = 7.6022\pm0.0007$~s, from a \emph{Suzaku} observation that was carried out on 2007 March 30 \citep{nakagawa09}, marking an average spin-down of $\dot{P}=(7.5\pm0.4)\times10^{-10}$~s s$^{-1}$ in the period from 2006 September 10 to 2007 March 30. In a more recent (unpublished) 
\xmm\ observation (obs.ID 0654230401, unpublished), performed on 2011 March 23, where we measured $P=7.7021\pm0.0001$~s. Comparing this value with the last published measurement of $P=7.6022$~s (2007 March 30, \citealt{nakagawa09}), it gives an average spin down rate of $\dot{P}=(7.95\pm0.06)\times10^{-10}$~s
s$^{-1}$ from 2007 March 30 to 2011 March 23. This is among the largest values measured so far for \sgr\ (see the McGill magnetar catalog, \citealt{olausen14}).
With these values at hand, we computed an upper
limit for the pulsed fraction of a periodic signal with $7.7 < P <
8.0$~s (assuming a spin down rate from zero to five times the last
observed $\dot{P}$). Taking into account the 154 independent
frequencies searched and following the recipes by \citet{israel96},
the 3$\sigma$ upper limit value is a poorly-constraining 64\% for a sinusoidal signal.

\section{Discussion}

The careful analysis of the X-ray emission of \sgr\ at $\sim$arcsecond scale shows that the comparison with the {\em Chandra}-HRC PSF strongly depends on the parameter {\tt Aspect\_Blur}, which accounts for the blurring of the image. The latter is a phenomenological way to account for the PSF features, and depends on the photon energy (i.e., on the source spectrum).

In order to better understand this issue, we have repeated the whole procedure for the point-like, relatively soft and bright X-ray Isolated NS RBS~1223. We analyzed the data of the HRC 87.2~ks observation of 2004 March 30, from which we extracted $\sim 4800$ counts in the central 2$"$ circular region around the source. We repeated the same procedure as for \sgr, in order to compare simulations and data. Again, we found that simulations with {\tt Aspect\_Blur=0.07} give too small PSFs compared with data, with an apparent $\sim 30\%$ counts excess in data. If instead {\tt Aspect\_Blur=0.15} is employed in the simulation, the PSF and data radial profiles are compatible. Since the value of the blur parameter is set only phenomenologically, a more accurate and definitive study at such small scales require more precise calibrations. Note also that the modeling of artifacts and blur at small scale has been included only in the latest version of {\tt MARX}, which has been a major revision. With previous versions of the software, the comparison of the radial profiles would have shown an excess in data, and could have been interpreted as diffuse emission. As a consequence, in order to draw physical interpretation from any HRC data at sub-arcsecond scale, we stress that the technical issues regarding simulated PSF should be carefully taken into account.

In conclusion, we do not see a statistically significant extended emission at small scales around \sgr, its observed emission is consistent with being point-like, both in the post-giant flare observation, and during quiescence. We do not detect any X-ray counterpart of the radio-nebula seen in 2005. On the other hand we found evidence in the 2005 data for diffuse emission on angular scale of about 10 arcseconds. Although a precise characterization of this diffuse emission is hampered by the poor statistics, by the large systematic errors due to the model-dependent PSF subtraction,  and by the lack of spectral information, we note that its size and intensity are consistent with a scattering halo caused by a dust cloud relatively close to the source.

\section*{Acknowledgments}
We acknowledge support from NewCompstar, and the grants AYA2012-39303 and SGR2009-811 (DV, NR). NR is supported by an NWO Vidi Award. We are grateful to: Lucien Kuiper and Wim Hermsen for their contribution in an earlier stage of the analysis of the 2005 data, Diego G\"otz for his useful comments, and Matteo Murgia for reanalyzing the VLA data for the time period following the giant flare.

\section*{References}

\bibliography{biblio}

\end{document}